\documentclass[aps,amsmath,amssymb, notitlepage, 
twocolumn,
nofootinbib,
]{revtex4-1}

\usepackage{amsmath,amssymb,bm,graphicx,dsfont,color}
\usepackage[latin9]{inputenc}
\usepackage{amsmath}
\usepackage{slashed}
\usepackage{dsfont}
\usepackage{amstext}
\usepackage{amssymb}
\usepackage{amsbsy}
\usepackage{amsthm}
\usepackage{epsfig}
\usepackage{graphicx}
\usepackage{bbm}
\usepackage{hyperref}
\usepackage{color}

\usepackage{slashed}

\renewcommand{\>}{\rangle}

\newcommand{\T}{\mathcal{T}}

\newcommand{\be}{\begin{equation}}
\newcommand{\ee}{\end{equation}}
\newcommand{\bea}{\begin{eqnarray}}
\newcommand{\eea}{\end{eqnarray}}

\newcommand{\LL}{\mathcal{L}}
\renewcommand{\vec}[1]{{\bf #1}}
\renewcommand{\hat}[1]{{\widehat #1}}

\begin{document}

\title{Composite fermi liquids in the lowest Landau level}
\author{Chong Wang}
\affiliation{Department of Physics, Harvard University,
Cambridge, MA 02138, USA}
\author{T. Senthil}
\affiliation{Department of Physics, Massachusetts Institute of Technology, Cambridge, MA 02139, U. S. A.}
\date{\today}
\begin{abstract}
{  We study composite fermi liquid (CFL) states in the lowest Landau level (LLL) limit at a generic filling $\nu = \frac{1}{n}$.  We begin with the old observation that, in compressible states,  the composite fermion in the lowest Landau level should be viewed as a charge-neutral particle carrying vorticity. This leads to the absence of a Chern-Simons term in the effective theory of the CFL.  We argue here that instead a Berry curvature should be enclosed by the fermi surface of composite fermions, with the total Berry phase fixed by the filling fraction $\phi_B=-2\pi\nu$. We illustrate this point with the CFL of fermions at filling fractions $\nu=1/2q$ and (single and two-component) bosons at $\nu=1/(2q+1)$. The Berry phase leads to sharp consequences in the transport properties including thermal and spin Hall conductances, which in the RPA approximation are distinct from the standard Halperin-Lee-Read predictions. We emphasize that these results only rely on the LLL limit, and do not require particle-hole symmetry, which is present microscopically only for fermions at $\nu=1/2$. Nevertheless, we show that the existing LLL theory of the composite fermi liquid for bosons at $\nu=1$ does have an emergent particle-hole symmetry. We interpret this particle-hole symmetry as a transformation between the empty state at $\nu=0$ and the boson integer quantum hall state at $\nu=2$.  {  This understanding enables us to define particle-hole conjugates of various bosonic quantum Hall states which we illustrate   with the  bosonic Jain and Pfaffian states.  For bosons at $\nu = 1$ we construct paired non-abelian states distinct from both the standard bosonic Pfaffian and it's particle hole conjugate and show how they may arise
naturally out of the neutral vortex composite fermi liquid. The bosonic }particle-hole symmetry can be realized exactly on the surface of a three-dimensional boson topological insulator.       We also show that with the particle-hole and spin $SU(2)$ rotation symmetries, there is no gapped topological phase for bosons at $\nu=1$.}
\end{abstract}

\maketitle
 
 \tableofcontents
 


\section{Introduction} 
Compressible metallic states of electronic systems at even denominator filling  in the quantum Hall regime were described in pioneering work\cite{hlr} by Halperin, Lee, and Read (HLR) as composite fermi liquids. The composite fermions\cite{jaincf} were obtained by attaching an even number of fictitious flux quanta to the electrons and move (after  a flux-smearing mean field approximation) in zero net magnetic field (see Ref.~\cite{lopez91} for application to incompressible states). This enables them to form a Fermi surface. In this construction the composite fermions carry the electric charge of the electron, and in addition the attached flux. Including fluctuations beyond the mean field leads to an effective action where the composite fermions are coupled minimally (due to their physical electric charge) to an external probe electromagnetic field, and to an internal $U(1)$ gauge field which has Chern-Simons dynamics, and which serves to implement the flux attachment. The HLR construction of composite fermi liquids leads (within an RPA treatment of the internal gauge fluctuations) to a number of predictions in experiments some of which have received striking confirmation (see, {\em eg}, Refs.~\cite{dassarmabook,willett97}). 

Despite these successes the HLR construction has some well appreciated problems\footnote{As emphasized recently\cite{balramjkj16}, these issues do not affect microscopic wavefunction based treatments of quantum Hall phenomena which work within the LLL}. In the limit that the typical interaction strength is much smaller than the cyclotron frequency, it should be appropriate to define the quantum hall problem purely in the Lowest Landau Level (LLL). However this limit of projection to the LLL is problematic for HLR. Formally, the flux attachment procedure involves all Landau levels: at the level of wave functions it involves multiplication by a phase factor $\Pi_{i < j} \left(\frac{z_i - z_j}{|z_i - z_j|}\right)^{2q}$ if $2q$ flux quanta are attached. The denominator clearly lives outside the LLL. The difficulty is also illustrated by the appearance of the bare electron mass $m_b$ in the composite fermion kinetic energy. The projection to the LLL involves taking the (apparently singular) limit $m_b \rightarrow 0$. 

In another pioneering work, Read\cite{read98} derived an alternate theory of a composite fermi liquid for the problem of bosons at filling $\nu = 1$.  This theory is manifestly in the LLL, but differs strikingly from the result of the HLR-RPA procedure applied to bosons at $\nu = 1$. Specifically the composite fermions in Read's theory are electrically neutral, and should instead be interpreted as vortices. A useful picture then is that of a quantum liquid of fermionic vortices.  These neutral vortices are not coupled minimally to the  external probe electromagnetic gauge field. However as is usual in dual vortex theories of boson liquids\cite{chandandual,mpafdhl89} they are coupled minimally to an internal non-compact $U(1)$ gauge field without a Chern-Simons term. The probe electromagnetic gauge field couples to the field strength of this internal gauge field (as is also common in dual vortex theories). 

A different problem - specific to the filling $\nu = 1/2$ - is that upon projecting to the LLL and restricting to, say, a two-body interaction there is a particle-hole symmetry which involves trading the electrons for holes obtained by removing electrons from a filled Landau level. Not being a LLL theory the HLR construction does not know anything about this symmetry. Recently a particle-hole symmetric theory for the $\nu = 1/2$ composite fermi liquid has been developed\cite{sonphcfl,dualdrcwts2015,dualdrmaxav,cfltislrev,geraedtsnum,msgmp15,kachru15} and takes a form distinct from the HLR-RPA action (even with renormalized parameters).  

The theory of the particle-hole symmetric $\nu = 1/2$ electronic composite fermi liquid however has striking similarities with  Read's LLL theory of bosonic $\nu = 1$ composite fermi liquids.  Indeed the particle-hole symmetric composite fermion is usefully understood as a strength-$4\pi$ vortex in the electronic fluid. This vortex is an electrically neutral fermion, and forms a Fermi surface.  It thus does not couple minimally to external electromagnetic fields but does couple minimally to an internal non-compact $U(1)$ gauge field without a Chern-Simons term. In contrast however to  the bosonic CFL at $\nu =1$,  in the  $\nu = 1/2$ particle-hole symmetric CFL when the composite fermion goes around the Fermi surface   it acquires a  Berry phase of $\pi$. 

In this paper we will bring out these similarities and differences between these quantum vortex liquid theories of composite fermi liquids. In addition to the two examples (electrons at $\nu = 1/2$ and bosons at $\nu = 1$) mentioned above we will consider more general fillings $\nu = \frac{1}{n}$, with even denominators for electrons(fermions) and odd denominators for bosons. We will first review, in Sec.~\ref{QVLT}, old expectations showing that a quantum vortex liquid theory is natural once the LLL restriction is imposed and  translation symmetry is  preserved.  A general feature common to such theories is the presence of an internal $U(1)$ gauge field but without a Chern-Simons term. We show instead that a new feature, a non-trivial Berry phase $\phi_B=-2\pi\nu=-2\pi/n$ appears on the composite fermion Fermi surface\footnote{A similar proposal has been made independently by Haldane which we became aware of while completing this paper.}. For the special case of electrons at $\nu = 1/2$, we recover the $\pi$ Berry phase of the particle-hole symmetric theory even though particle-hole symmetry per se plays no direct role in our arguments. For bosons at $\nu=1$ we also recover Read's theory. An especially interesting case is two-component bosons with full $SU(2)$ spin rotation symmetry: we will show in Sec.~\ref{upspin} that the composite fermions again form Dirac fermi seas.

The Berry phase on the fermi surface comes with real physical consequences. In Sec.~\ref{transport}, we discuss the effect of Berry phase on transport properties, focusing on thermal Hall conductance and spin Hall conductance (in cases with unpolarized spins). We will see that the Berry phase leads to predictions that are distinct from those in standard HLR-RPA theory.

We emphasize that our theory relies on the LLL limit only, and does not require particle-hole symmetry to be present. In fact particle-hole symmetry only exists microscopically for electrons at $\nu=1/2$. However, we will show in Sec.~\ref{PHS} that for bosons at $\nu=1$, which does not possess particle-hole symmetry microscopically, there is a low-energy emergent particle-hole symmetry in the quantum vortex liquid phase described by Read's theory. {This possibility was raised in the Hamiltonian approach in Ref.~\cite{msgmp15}.} We show that this emergent particle-hole symmetry is related to a particle-hole transformation of bosonic quantum hall states: one that transforms a state at filling $\nu$ to a state at filling $2-\nu$. We also discuss particle-hole conjugates of bosonic Jain states and Pfaffian states.  In particular a bosonic anti-Pfaffian state at $\nu = 1$ is naturally obtained through the particle-hole transformation and is distinct from the bosonic Pfaffian state at the same filling.  We show that angular momentum $l = \pm 1$ pairing of the neutral vortex composite fermions in Read's theory leads to non-abelian topological ordered states that are distinct both from the standard bosonic Pfaffian and the anti-Pfaffian we introduce. This difference is a manifestation of the Fermi surface Berry phase of $-2\pi$ that is associated with these neutral composite fermions.  We discuss the relation between these various paired states and show how they can all be obtained from pairing either starting from Read's theory or the original HLR theory. We show in Sec.~\ref{3dSPT} how these observations about Read's theory find a natural ``home" at the surface of a three dimensional  bosonic topological insulator in close analogy to points of view that have proven powerful for the electronic half-filled Landau level. In Sec.~\ref{SEG} we discuss a peculiar feature of   two-component bosons at $\nu=1$: with full $SU(2)$ rotation and particle-hole symmetry, the system cannot be in an incompressible, topologically ordered state. Though perhaps not very  pertinent to quantum Hall systems, this result  is of primary interest in the context of three dimensional Symmetry Protected Topological (SPT) phases. It provides an example of a bosonic SPT phase which does not admit a symmetry preserving gapped surface. Such `Symmetry enforced gaplessness" was previously described for a fermionic topological superconductor in Ref.~\cite{3dfSPT2}. 

\section{HLR and the Lowest Landau Level limit}
\label{hlrlll}
We first review some old ideas on the HLR theory and it's fate in the Lowest Landau Level (LLL) that will provide the background we will build on in this paper.  

We start by considering the HLR approach\cite{hlr} to compressible quantum Hall states at filling $\nu=1/n$ (with $n$ even/odd depending on whether the microscopic particles are fermions/bosons). This approach 
 begins with an exact flux attachment transformation which converts electrons to composite fermions $\psi_c$.  The corresponding action takes the form\footnote{Very strictly speaking, instead of having a fractional level Chern-Simons term $\frac{1}{4\pi n}a'da'$, it is more well defined to introduce another gauge field $\alpha$ and replace the Chern-Simons term by $\frac{1}{2\pi}a'd\alpha-\frac{n}{4\pi}\alpha d\alpha$. The gauge fields in the latter formulation have standard flux quantization and therefore easier to manipulate. However, for our purpose it suffices to integrate out $\alpha$ and work with the form used in the main text.}  
\bea
\label{HLR}
\mathcal{L}_{HLR}=&&\mathcal{L}[\psi_c,a'_{\mu} + A^{tot}_{\mu}]+\frac{1}{4n\pi}\epsilon_{\mu\nu\lambda}a'_{\mu}\partial_{\nu}a'_{\lambda} 
\eea
Here $A^{tot} = A^{bg} + A$ is the total external gauge field which includes both an $A^{bg}$ associated with the background magnetic field and a  `probe'  $A$.  The internal `Chern-Simons' gauge field $a'$ implements the flux attachment . The theory proceeds by making a mean field approximation where the $a'$ acquires an expectation value that cancels the background $A^{bg}$. The composite fermions then move in zero net field and can form a Fermi surface. Fluctuations beyond the mean field lead to a field theory of these composite fermions coupled minimally to both $a'$ and to the probe field $A$. A low energy effective theory is obtained by truncating to $\psi_c$ modes that live near the Fermi surface. This theory then takes the form, after the shift $a' + A = a$, 
\bea
\label{RPA}
\mathcal{L}_{RPA}=&&\mathcal{L}_{FS}[\psi_c,a_{\mu}]+\frac{1}{4n\pi}\epsilon_{\mu\nu\lambda}a_{\mu}\partial_{\nu}a_{\lambda}-\frac{1}{2n\pi}\epsilon_{\mu\nu\lambda}a_{\mu}\partial_{\nu}A_{\lambda} \nonumber \\ &&+\frac{1}{4n\pi}\epsilon_{\mu\nu\lambda}A_{\mu}\partial_{\nu}A_{\lambda}+..,
\eea
where $\mathcal{L}_{RPA}[\psi_{c},a_{\mu}]$ describes fermions $\psi_c$ {  modes restricted to be near the fermi surface} minimally coupled with the emergent gauge field $a_{\mu}$. 
Specific predictions for a number of physical quantities follow from an RPA treatment of the internal gauge fluctuations. We emphasize that even though Eq.~\eqref{HLR} and \eqref{RPA} take very similar forms, they bear very different physical meanings: the former is an exact reformulation of the original problem which is still strongly coupled, and the latter is a possible effective field theory that describes the low energy physics. 

The standard criticism of HLR theory in the LLL limit is that the limit of zero bare electron mass appears singular in the reformulation of Eq.~\eqref{HLR}. However, this on its own does not rule out the effective field theory Eq.~\eqref{RPA} as a candidate theory of compressible states, since one expects the parameters in the effective theory, including the fermion mass, to be renormalized from the bare values. Optimistically therefore we can suppose that the action in Eq.~\eqref{RPA} still works in the LLL limit but with (a priori unknown) effective parameters. 

The more serious problem with Eq.~\eqref{RPA} in the LLL limit comes from the electric charge carried by the composite fermions. Taking the equation of motion $\frac{\delta\mathcal{L}_{RPA}}{\delta a_{\mu}}=0$, it is obvious that the $\psi_c$ fermions in the RPA theory carries the electric charge $q_A=1$. However, it was pointed out by Read\cite{read94} and subsequently others\cite{rsgm97,Pasquier1998,dhleephcf98,sternetal99} that the composite fermion in LLL, at the compressible filling fractions $\nu=1/n$, should be charge-neutral. 
As this observation will play a crucial role in the present paper we now review the reasoning of Ref. \cite{read94}. 

Thinking in terms of wavefunctions, flux attachment introduces a factor $\frac{\left(z_i - z_j \right)^{n}}{\left|z_i-z_j\right|^n}$ which is not holomorphic, and hence not in the LLL.  A better alternative\cite{jaincf} is to do vortex attachment which introduces a multiplicative factor $\left(z_i - z_j \right)^{n}$ without the non-holomorphic denominators. Note that the vortex attachment forces a change in the amplitude of the wavefunction by forcing zeroes of the wave function in the vicinity of the original particles. Thus the vortices should be viewed as correlation holes attached to electrons. In this procedure a composite fermion is viewed as the original particle  bound to a strength-$n$ vortex (at filling $1/n$). This composite fermion will then have its electric charge reduced by the charge of this vortex/correlation hole. 

By how much is the charge depleted at a vortex? Consider taking a single $2\pi$ vortex around a loop of area ${\cal A}$.  During this process the vortex will pick up a Berry phase determined by the background charge density is uniform as is appropriate for a translation invariant system.  At a filling fraction $\nu$ this phase is  $-2\pi \rho_e {\cal A} = -\nu B {\cal A}$ which is precisely the phase acquired by a charge of strength $-\nu$ moving in the magnetic field.  For $\nu = \frac{1}{n}$ the charge of an $n$-fold vortex is then $-1$.  This exactly compensates for the charge of the original particle. The composite fermion formed by binding the original particle to an $n$-fold vortex is thus expected to be neutral. 

As a check, this argument can be extended to the Jain sequence of plateaus at $\nu = \frac{p}{np+1}$ proximate to the compressible state at $\nu = \frac{1}{n}$. (Here $p$ is an integer). The bound state of the original particle and an $n$-fold vortex will then have charge $1 - n\nu = \frac{1}{np+1}$. It is easy to also check that it has exchange statistics $\theta = \frac{\pi np}{np+1}$. As we approach the compressible state at $\nu = \frac{1}{n}$ through a sequence of incompressible states at $\nu = \frac{p}{np+1}$ by taking the limit $p \rightarrow \infty$ we see that the charge of the particle-$n$-fold vortex bound state goes to zero and its statistical angle goes to $\pi$. 

Thus we anticipate that a purely LLL theory of the compressible states will be formulated in terms of composite fermion fields that are electrically neutral rather than in terms of the charged composite fermions of the original HLR theory. 

 The charge-neutrality of composite fermions disfavors the simple HLR effective theory as a description of the compressible states in LLL. We should emphasize here that the difficulty of the HLR  theory in LLL has nothing to do with its non-fermi liquid nature and various divergencies of quantities at low energy that are hard to control. In fact, one can imagine having a very long-range interaction $V\sim 1/r^{1-\epsilon}$, in which case the HLR effective theory flows to  a simple fermi liquid fixed point. The charge-neutrality issue still remains in this case, and the LLL problem should presumably not be able to flow to the same fixed point.
 
  Though the LLL composite fermions are charge neutral, they carry non-zero vorticity. Thus the theory of the compressible state should take the form of a `dual' vortex theory - as in the familiar dual descriptions of bosons in zero field\cite{chandandual,mpafdhl89}, the vortex degrees of freedom will be coupled minimally to an internal non-compact $U(1)$ gauge field whose 3-flux is the physical 3-current of the underlying particles. We will call this the quantum vortex liquid theory.  
 
 To get an appreciation of how the charge neutrality condition may be implemented in such a theory, let us first understand better the HLR effective theory. Varying the HLR action with respect to $a_\mu$ we see that the composite fermion current $j_\mu$ satisfies
 \begin{equation} 
j_\mu =- \frac{1}{4\pi} \epsilon_{\mu\nu\lambda} \partial_\nu \left(a_\lambda - A_\lambda \right)
\end{equation}
However the physical electrical current (denoted $J_\mu$) is also given by the same expression. This is exactly as expected for the HLR composite fermion which carry the full charge of the electron.   Note the crucial role played by the Chern-Simons term for the internal gauge field. If just this term were absent from the action we would not have been able to identify the physical and composite fermion currents. 

Thus we expect that any putative quantum vortex liquid theory of the compressible states in the LLL will have a composite fermion coupled minimally to an internal $U(1)$ gauge field $a_\mu$ but without a Chern-Simons term.  The 3-flux of $a_\mu$ is the physical 3-current of the underlying particles. 

A final consideration is that though the composite fermions are electrically neutral they are expected to carry a dipole moment proportional to their momentum. This dipole moment will point perpendicular to the direction of the momentum. This dipole moment comes from a displacement of the vortices away from the position of the electron, as suggested by heuristic wavefunction arguments\cite{read94}.  Indeed, it is a general restriction of the LLL that the total dipole moment operator $\vec D$  in a many body state has an exact relationship with the total momentum operator $\vec P$. They satisfy $\vec D = l_B^2   \hat {z} \times \vec P$ where $l_B = \sqrt{\frac{1}{B}}$ is the magnetic length. It is natural that this many body constraint is implemented at the level of individual composite fermions in the theory.

What we have reviewed thus far are ideas from the 1990s on the shape of a purely LLL theory of compressible composite fermi liquid states in terms of neutral dipolar composite fermions. Below we will describe a specific effective theory which realizes these hopes. We will see that it involves a new ingredient - a Fermi surface Berry phase.

\section{Quantum vortex liquid theory of compressible quantum hall states}
\label{QVLT}

Microscopically  the problem of describing states in the LLL should be formulated as follows. First specifiy the Hilbert space of many particle states (whether particles are bosons or fermions, and at what filling).  Then work with a  Hamiltonian described in terms of projected density operators that satisfy the Girvin-MacDonald-Platzmann algebra\cite{GMP}.  Going from this microscopic formulation to a low energy effective theory of the compressible states is a rather complex task. To this date it has only been taken to completion\cite{read98} in the case of bosons at $\nu = 1$.  We will not attack this problem head-on in this paper. 
We instead ask a simpler question: what effective theory, presumably fermi-liquid-like, can emerge from Eq.~\eqref{HLR} while satisfying the charge-neutrality of composite fermions which nevertheless carry vorticity?

\begin{figure}
\begin{center}
\includegraphics[width=3in]{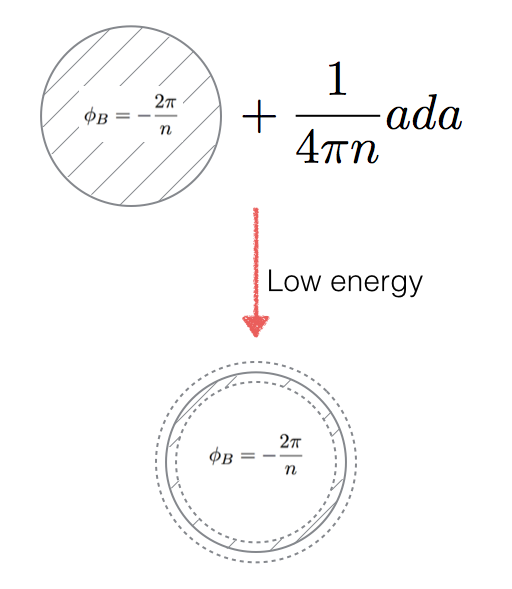}
\end{center}

\caption{Cancelation of the Chern-Simons term by integrating out composite fermions deep in the fermi sea. Final theory: a composite fermi surface with Berry phase $\phi_B=-2\pi/n$. }
\label{flow}

\end{figure}

 The answer is almost obvious in hindsight: we postulate a fermi surface formed by $\psi_c$, with a Berry phase on the fermi surface $\phi_B=-\frac{2\pi}{n}$.\footnote{A concrete field theory realization of a fermi surface Berry phase $\phi_B$ would be a massive Dirac fermion, with chemical potential $\mu$ fine-tuned so that the fermi surface covers exactly a Berry phase of $\phi_B$. This, however, should be viewed only as an alternative UV-completion of our low-energy theory. The actual microscopic physics in the LLL may not involve the Dirac nature in any meaningful way.} As we go to low energy by integrating out fermions deep in the fermi sea, another Chern-Simons term appears due to the anomalous Hall conductance from the fermions\cite{HaldaneBerry}. The coefficient of this new Chern-Simons term is $\frac{\phi_B}{8\pi^2}=-\frac{1}{4n\pi}$, which exactly cancels the original Chern-Simons term in Eq.~\eqref{HLR} (see Fig.~\ref{flow}). The final effective theory has no Chern-Simons term for $a_{\mu}$, and the composite fermions are charge-neutral. We will denote these charge-neutral composite fermions $\psi_v$ to distinguish them from the HLR composite fermions $\psi_c$.

The effective Lagrangian for these neutral composite fermions $\psi_v$ will take the form 
\begin{equation}
\label{vortex}
{\cal L} = {\cal L}_{\phi_B}[\psi_v, a_\mu] - \frac{1}{2n\pi } \epsilon_{\mu\nu \lambda} a_\mu \partial_\nu A_\lambda +  \frac{1}{4n\pi }\epsilon_{\mu \nu \lambda} A_\mu \partial_\nu A_\lambda.
\end{equation}
Here the first term describes the composite fermions near their Fermi surface coupled minimally with the internal non-compact $U(1)$ gauge field $a_{\mu}$. Since these are strength $n$ vortices, they couple to the external probe gauge fields $A_{\mu}$ as indicated in the second term. The last term is a background Hall conductivity for the probe gauge field which ensures that the theory really arises in a strictly two dimensional system. 
Most crucially the composite fermions see a Berry phase of $\phi_B=-\frac{2\pi}{n}$ when they go around their Fermi surface. The Lagrangian above supplemented with this Berry phase is the proposed quantum vortex liquid description of the composite fermi liquid of electrons/bosons at filling $1/{n}$ in the LLL. 

We emphasize again that we did not derive the theory in Eq.~\eqref{vortex} from the microscopic  LLL problem. Rather, we postulated the low energy effective theory starting from HLR formulation in Eq.~\eqref{HLR}, and argued that this theory, unlike the standard HLR-RPA theory, is compatible with the physical requirement of charge-neutral composite fermions that arises in the LLL.

Strong support for our proposed quantum vortex liquid theory  is provided by studying some consequences of the proposed Lagrangian. Since these composite fermions carry $2\pi n$ vorticity but are electrically neutral,  it follows that their density $n_v$ is determined entirely by the external magnetic field: 
 \begin{equation}
 \label{cfrho}
n_v = \frac{B}{2\pi n}
\end{equation}
It is instructive to consider what happens when we move slightly away from $\nu = \frac{1}{n}$. The composite fermions feel a magnetic field 
\be
\label{fluxcharge}
B^*=B-2\pi n\rho_e,
\ee
where $\rho_e$ is the electron/boson density. Let us now ask about a `measurement' of the Fermi surface Berry phase by tracking the resistivity minima plotted as a function of $1/B^*$ at fixed $n_v$, {\em i.e} at fixed $B$. It is well known that these occur at a Jain sequence with filling $\nu = \frac{p}{np+ 1}$. This corresponds to effective fields that satisfy
\begin{equation}
\label{shift}
\frac{1}{B^*} = \frac{p + \frac{1}{n}}{\frac{B}{n}}
\end{equation}
The shift from an integer in the numerator is $-\frac{\phi_B}{2\pi}$ where $\phi_B$ is the Fermi surface Berry phase (see Ref.~\cite{Roth, Mikitik}). It follows that this Berry phase is $-\frac{2\pi}{n}$ exactly as proposed in the effective field theory Eq.~\eqref{vortex}. 

We should point out that the shift from integer values in Eq.~\eqref{shift} only applies when the fermi surface has a Berry phase, but does not contribute to the overall Hall conductance. For free fermions this happens in special cases, such as on the surface of a topological insulator, or in graphene when the quantity is divided by four to obtain the value from each Dirac cone. In our case the ``missing" hall conductance of composite fermion simply comes from the Chern-Simons term due to flux-attachment. These issues are discussed in more detail in Appendix \ref{fsb}.

To obtain further insight into this Berry phase  let us first develop a heuristic physical picture of the composite fermion that generalizes the considerations of Ref. \cite{cfltislrev} from $\nu = \frac{1}{2}$ to general $\nu = \frac{1}{n}$.  The composite fermion is built by attaching $n$ vortices to the underlying particles. We begin by first considering the case where the particles are fermion (and hence $n = 2q$ is even). Of these $n$ vortices the antisymmetry of LLL fermion wave functions forces one (or more generally an odd number) vortex to be exactly on top of the electron. The remaining $n- 1$ vortices will be displaced away from the electron in the direction perpendicular to the composite fermion momentum. The composite fermion at general $n$ is thus a neutral dipole with each end of the dipole carrying electric charge $\pm \left(1 - \frac{1}{n} \right)$. When one end of the dipole is rotated fully about the other end by an angle $2\pi$ there is a phase $\frac{-2\pi}{n}$. If these fermions form a Fermi surface, then as a composite fermion moves a full circle around this surface the momentum, and hence the internal dipole moment, rotate by $2\pi$. It follows that there is a Fermi surface Berry phase of   $-\frac{2\pi}{n}$.  

Is there an analagous heuristic picture for the composite fermion $\psi_v$ when the underlying particles are bosons? Now the symmetry of the wavefunction does not force any vortex to sit exactly on top of the particle. Nevertheless we are free to put any even number of vortices on the particle, and displace the remaining away.  Consider the case where $n-1$ vortices sit on the particle, and the remaining vortex is displaced away. The end with $n-1$ vortices then has charge $\frac{1}{n}$ and the other end has charge $-\frac{1}{n}$. The Berry phase picked up when one end goes around the other is precisely $- \frac{2\pi}{n}$. Forming a Fermi surface of these particles we will then get a Berry phase of $-\frac{2\pi}{n}$. 

The picture of $n-1$ vortices sitting exactly on top of the particle while the remaining one is displaced away actually works for any $n$ ({\em i.e}, whether the microscopic particles are fermions or bosons), and gives a Berry phase of $-\frac{2\pi}{n}$. Why make this choice as compared to say some other way of distributing the $n$ vortices? We do  not have a satisfactory answer to this question. However the heuristic picture sketched above does, we believe, help develop some intuition about the origin of the Fermi surface Berry phase. 
This picture also suggests that for any $n$, a useful wavefunction for the CFL  is given by 
\begin{equation}
\psi(z_1, z_2,........z_N) = \Pi_{j< i} \left(z_i - z_j \right)^{n-1} \psi^b_{\nu = 1} (z_1, ......., z_N)
\end{equation}
where $\psi^b_{\nu = 1}$ is the wavefunction of the bosonic CFL at filling factor $\nu = 1$.

\subsection{Fermions at $\nu=1/2$}

There are two special cases in which theories of the form of Eq.~\eqref{vortex} are more solidly justified. The first is fermions at $\nu=1/2$. It is well known that fermions at $\nu=1/2$ in the LLL limit have an emergent particle-hole symmetry ($C$), which does not seem to be preserved by the standard RPA effective field theory\cite{klkgphhlr}.  This was emphasized recently\cite{Barkeshli15} by the construction of a particle-hole conjugate to the standard HLR theory which leads to a seemingly distinct effective field theory. Another effective field theory describing a compressible state was proposed recently by Son\cite{sonphcfl}, which manifestly preserves the particle-hole symmetry. The low energy theory takes exactly the same form as Eq.~\eqref{vortex}, with the Berry phase $\phi_B=\pi$. A fermi surface with $\pi$-Berry phase can be conveniently represented as a Dirac fermion:
\bea
\LL[\psi,a_{\mu}]=&&\bar{\psi}(i\slashed{\partial}+\slashed{a}+\mu\gamma^0)\psi \nonumber \\
&&-\frac{1}{4\pi}\epsilon^{\mu\nu\lambda}a_{\mu}\partial_{\nu}A_{\lambda}+\frac{1}{8\pi}\epsilon^{\mu\nu\lambda}A_{\mu}\partial_{\nu}A_{\lambda}
\eea
and the particle-hole symmetry $C$ is represented as an anti-unitary symmetry that acts like time-reversal on the Dirac composite fermions
\be
\psi_c\to i\sigma^y\psi_c.
\ee
We emphasize that the ``Diracness" is manifested through the $\pi$-Berry phase, rather than the Dirac cone itself, which is far from the fermi surface and has no real physical meaning at low energy. Indeed a recent paper by Balram and Jain\cite{balramjkj16} argues (through calculations on wavefunctions for proximate incompressible states) for lack of evidence of the linear dispersion 
associated with the Dirac cone. 

The Dirac composite fermi liquid state was justified through a surprising charge-vortex duality in $(2+1)d$ between free Dirac fermion and quantum electrodynamics\cite{dualdrcwts2015,dualdrmaxav,cfltislrev,dualdrMAM}. It could also be understood as a critical point separating HLR and anti-HLR states\cite{Barkeshli15,Mulligan16}. The $\pi$-Berry phase of the composite fermions has been verified numerically\cite{geraedtsnum} through the absence of $2k_f$ scattering peaks for particle-hole symmetric correlators. In the full theory (without the LLL restriction), the $\pi$-Berry phase can change (to trade a Chern-Simons term) by mixing the upper and lower Dirac band of the composite fermions, if particle-hole symmetry is absent. However,  in a strictly LLL theory, we now argue that the $\pi$ Berry phase survives even if particle-hole symmetry is not present in the microscopic Hamiltonian. 

Consider starting initially with $C$ symmetry present, and go to the low energy theory near the Fermi surface of the Dirac composite fermion. Now break the $C$ symmetry explicitly through some weak perturbations. Naively, in the absence of $C$ we can add a mass term $\bar{\psi_v}\psi_v$ to the low energy Dirac action. This will force the Dirac `spin' to tilt out of the $xy$ plane. When the resulting fermion goes around the Fermi surface the Berry phase will then deviate away from $\pi$, if such a mass term were allowed. However in the LLL we can argue that such a mass term is prohibited. The key is the understanding developed in Ref. \cite{cfltislrev} of the $x,y$ components of the Dirac spin as the composite fermion dipole moment. A rigid restriction of the LLL limit is that this dipole moment is rotated from the composite fermion momentum by an angle of $\frac{\pi}{2}$ in the anticlockwise direction. 
In the LLL the other polarization of the composite fermion, where the dipole moment (and hence the $x,y$ components of Dirac spin) are rotated by $-\frac{\pi}{2}$ from the momentum direction are simply not there in the Hilbert space. In other words, it is impossible in a purely LLL theory to flip the direction of the dipole moment of the composite fermion while preserving its momentum. Equivalently we can say that the negative energy Dirac sea does not really exist in a purely LLL theory. 

Now the Dirac mass operator $\bar{\psi_v}\psi_v$  has the precise effect of mixing the two components of the Dirac spin at a fixed momentum. But since only one component exists as a physical state in the LLL this mixing term is not allowed. The inability to add this mass term in the LLL means that the $\pi$ Berry phase is stable to breaking $C$ symmetry.  

Thus the LLL protects the $\pi$ Berry phase even in the absence of exact particle-hole symmetry. 
 Rather, particle-hole symmetry is an additional feature that, if present, can be captured by the quantum vortex theory. This is quite different from previous views on the half-filled Landau level problem.
 
 Note however that, in the absence of particle-hole symmetry, the effects of the $\pi$ Berry phase will not be readily visible through the suppression of $2K_f$ singularities.  These singularities will be suppressed in the correlators of the density of composite fermions. But it is hard to know what microscopic electron operators couple to the composite fermion density in the absence of particle-hole symmetry.  


\subsection{Bosons at $\nu=1$}

Another example, developed even earlier, considers bosons at $\nu=1$. It was realized by Read\cite{read98} through an elaborate derivation that in the LLL, the compressible state of bosons at $\nu=1$ should be described by a simple fermi surface coupled with the emergent gauge field $a_{\mu}$, without a Chern-Simons term for $a_{\mu}$. 

The general theory in Eq.~\eqref{vortex}, when applied to $\nu=1$, gives the fermi surface a total Berry phase $\phi_B=-2\pi$, which would not be dynamically observable. Therefore it agrees well with Read's theory, which was constructed microscopically. 

Unlike fermions at $\nu=1/2$, a system of bosons at $\nu=1$ does not have particle-hole symmetry microscopically. As we will see in Sec.~\ref{PHS}, a particle-hole symmetry emerges nevertheless at low energy in Read's theory.  Such a possibility was raised recently in Ref.~\cite{msgmp15}. This further emphasizes the point that LLL descriptions of composite fermi liquids naturally  lead to quantum vortex liquid theories with associated Fermi surface Berry phases. Particle-hole symmetry is not a prerequisite though the quantum vortex liquid theory is fully capable of incorporating it. 

\subsection{Compressible states with unpolarized spins}
\label{upspin}

It will be very illuminating to consider electrons/bosons with unpolarized spins. In general, when the spin is not fully polarized (when the Zeeman coupling is weak), we expect two fermi surfaces formed by the composite fermions. In the quantum vortex theory Eq.~\eqref{vortex}, this requires the total Berry phase $\phi_B=-2\pi\nu$ to be distributed in some way among the two fermi surfaces. A special situation is when the system possesses the full $SU(2)$ spin-rotation invariance, in which case the total Berry phase must be divided evenly between the two fermi surfaces. This gives a Berry phase $\phi'_B=-\pi\nu$ for each fermi surface.

A very interesting special case is when $\nu=1$. With single-component boson (fully polarized spins), the total Berry phase is $\phi_B=-2\pi$, which does not seem to have any nontrivial consequence. However, with full spin $SU(2)$ symmetry, the Berry phase on each fermi surface becomes $\phi'_B=-\pi$, which appears to be Dirac-like! Again we can conveniently represent the theory as a quantum-electrodynamics with two Dirac fermions:
 \bea
\label{DCFL}
\LL[\psi_{\alpha},a_{\mu}]&=&\sum_{\alpha=\uparrow,\downarrow}\bar{\psi}_{\alpha}(i\slashed{\partial}+\slashed{a}+\mu\gamma^0)\psi_{\alpha} \nonumber \\
&-&\frac{1}{2\pi}\epsilon^{\mu\nu\lambda}a_{\mu}\partial_{\nu}A_{\lambda}+\frac{1}{4\pi}\epsilon^{\mu\nu\lambda}A_{\mu}\partial_{\nu}A_{\lambda},
\eea
where the emergent $U(1)$ gauge field $a_{\mu}$ is coupled to two flavors of two-component Dirac fermions $\psi_{\alpha}$, one for each spin. We choose the gamma matrices to be $\gamma^0=i\tau^2,\gamma^1=\tau^3,\gamma^2=\tau^1$, where $\tau^i$ are Pauli matrices in the Dirac pseudo-spin space. The Dirac fermions are at finite chemical potential $\mu$. $A_{\mu}$ denotes the probe (non-dynamical) gauge field that couples with the physical charge current. {Two-component bosons at total filling $\nu=1$ were studied numerically in Ref.~\cite{WuJain}, and a fermi-liquid-like state was found. It will be interesting to see if any signature of the Berry phase proposed here can be detected numerically.}

One can actually understand the necessity of Dirac fermions here using familiar results from field theory literature: it is well known\cite{Kapustin} that the monopole operator for Eq.~\eqref{DCFL} -- one that changes $a_{\mu}$-flux by $2\pi$ -- carries $SU(2)$ spin-$1/2$. Now the monopole operator is a local operator that carries physical charge one (see Eq.~\eqref{fluxcharge}), so it must corresponds to the physical boson operator. This means the physical boson must carry spin-$1/2$ -- exactly what we required. If we used simple fermi surfaces (without Berry phase) instead, the physical boson -- the monopole -- would carry no spin. 

We emphasize the importance of the full spin $SU(2)$ symmetry (or at least a $U(1)\rtimes\mathbb{Z}_2$ subgroup). If only the $S_z$ component is conserved, we can have a state in which all the $-2\pi$ Berry phase is enclosed within one fermi surface while the other fermi surface has no Berry phase. This would look identical to the original Read's theory with two simple Fermi surfaces.

\section{Transport signatures of the Berry phase}
\label{transport}

We now discuss physical consequences of the Berry phase in terms of transport properties. {  First consider electrical transport. The electrical conductivity tensor $\sigma$ is the sum of two contributions. The background Chern-Simons term for the probe field $A_\mu$  in the Lagrangian lead to a background Hall conductivity $\sigma_{xy}^{bg} = \frac{e^2}{nh}$ which will add to the conductivity tensor $\sigma^*$ coming from the composite fermion/vortex liquid. Thus 
\begin{equation}
\sigma_{ij} = \sigma^*_{ij} + \frac{e^2}{nh} \epsilon_{ij}
\end{equation}
Here $\epsilon_{ij}$ is antisymmetric and $\epsilon_{xy} = 1$. The composite fermion contribution is readily seen, from the vortex liquid interpretation,  to be 
\begin{equation}
\sigma^*_{ij} = \delta_{ij} \frac{e^4}{(nh)^2 \sigma_v}
\end{equation}
where $\sigma_v$ is the RPA expression for the conductivity of the composite fermions ({\em i.e} the vortex conductivity).   As a function of wavenumber $q$, the composite fermion conductivity $\sigma_v$ takes the well-known form 
\begin{eqnarray}
\sigma_v & = & \frac{e^2K_F l}{4\pi \hbar}, ~~ q \ll \frac{2}{l} \\
& = & \frac{e^2 K_F}{2\pi \hbar q}, ~~ q \gg \frac{2}{l}
\end{eqnarray}
where $l$ is the impurity induced mean free path for the composite fermions. The final answer for the measured longitudinal conductivity is vey similar to that within the standard HLR-RPA theory. 

Following the reasoning of Ref.~\cite{cfltislrev}, the longitudinal thermal conductivity will be metallic but will have a dramatic violation of the Wiedemann-Franz law.  This too is a feature of both the vortex and HLR theories.

 However,  thermoelectric properties will be different between the two theories at the RPA level: it was pointed out in Ref.~\cite{PSV}, in the context of Dirac composite fermions for $\nu=1/2$, that the charge-neutrality of the composite fermion leads to a nonzero Nernst effect. The same argument applies to any theory of the form Eq.~\eqref{vortex}, and leads to a nonzero Nernst effect. For the HLR-RPA theory in Eq.~\eqref{RPA}, the Nernst coefficient vanishes at the RPA level\cite{PSV}, but (similarly to the thermal Hall effect discussed below) a nonzero value is expected beyond the RPA approximation, even though it is not clear to us how to calculate the correction quantitatively.

Below we consider two additional effects that are manifestly related to the existence of the Berry phase: the first is the thermal hall conductance, and the second is the spin hall conductance (when spins are unpolarized).

\subsection{Thermal Hall conductance}
\label{thermalhall}

It is well known that a Berry phase enclosed by a fermi surface gives not only electric hall conductance, but also thermal hall conductance. We now calculate the thermal hall conductance $\kappa_{xy}$ of the vortex theory Eq.~\eqref{vortex} using the standard RPA (Ioffe-Larkin) approach. 

The calculation can be best understood using a slave-particle (parton) representation of the theory. We decompose the physical electron into two particles:
\be
c=bf,
\ee
where $b$ is a boson and $f$ is a fermion.\footnote{If the physical particle $\mathcal{C}$ is a boson, the slave particle $b$ would be fermion. The rest of the argument will be unchanged.} Flux-attachment corresponds to putting $b$ into a Laughlin state at $\tilde{\nu}=\nu$. In the quantum vortex theory Eq.~\eqref{vortex}, the $f$ fermion encloses a Berry phase $\phi_B=-2\pi\nu$. The Ioffe-Larkin composition rule requires the thermal hall conductance be the the sum of the two slave particles:
\be
\label{IoffeLarkin}
\kappa_{xy}=\kappa_{xy}^b+\kappa_{xy}^f=(1-\nu)\frac{\pi^2k_B^2T}{3h}.
\ee

 Here we comment on the validity of Ioffe-Larkin approximation on thermal hall conductance. Unlike for charge resistance, Ioffe-Larkin is usually not well justified for thermal conductance. For example, for incompressible state at $\nu=1/3$, Ioffe-Larkin gives $\kappa_{xy}=\kappa_{xy}^b+\kappa_{xy}^f=2$ which is wrong. This is because the gauge field $a_{\mu}$ gives another contribution to thermal transport which cannot be neglected. Likewise for HLR-RPA theory in Eq.~\eqref{RPA}, the low-energy dynamics of the gauge field has a Chern-Simons term, which is expected to contribute to $\kappa_{xy}$ even though it is hard to calculate quantitatively. Therefore the naive Ioffe-Larkin result for the HLR-RPA theory, which would give $\kappa_{xy}=\frac{\pi^2k_B^2T}{3h}$ for any CFL state regardless of filling, cannot be trusted. However, for the quantum vortex theory in Eq.~\eqref{vortex}, the low-energy dynamics of the gauge field, by design, has no Chern-Simons term. Thus we expect the gauge field to contribute only to the diagonal part of thermal hall conductance (assuming higher derivative terms are irrelevant). The vortex theory in Eq.~\eqref{vortex} is a very special case in which Ioffe-Larkin is justified for calculating $\kappa_{xy}$ and the result in Eq.~\eqref{IoffeLarkin} can be trusted. 


The case with $\nu=1$ bosons is especially interesting here: the vortex theory predicts that $\kappa_{xy}=0$ while the standard RPA theory gives nonzero $\kappa_{xy}$. As we will see in Sec.~\ref{PHS}, this is related to the emergent particle-hole symmetry for $\nu=1$. 

For fermions at $\nu=1/2$, the vortex theory predicts $\kappa_{xy}=\frac{1}{2}\frac{\pi^2k_B^2T}{3h}$ which is consistent with the particle-hole symmetry in the lowest Landau level.

\subsection{Spin Hall conductance}
\label{spinhall}

We now discuss spin hall conductance with unpolarized spins. We should first clarify what we mean by {  spin} hall conductance here: we are not referring to spin-charge hall conductance, meaning a spin current induced by a transverse electric field. Rather we are referring to spin-spin hall conductance, meaning a spin current induced by a transverse gradient of Zeeman energy. Formally the spin-spin hall conductance is represented as a Chern-Simons term of the gauge field that couples to the spin degree of freedom, while the spin-charge hall conductance is a mutual Chern-Simons term between the spin and charge gauge field. The latter makes sense only if the spin rotation group is $U(1)$, with only $S_z$ conservation; while the former makes sense even if we have the full $SU(2)$ symmetry.

The difference between the vortex theory and the standard {   HLR-RPA} theory is sharpest when we have full $SU(2)$ symmetry, and we expect the rest of the analysis to hold when the breaking from $SU(2)$ to $U(1)$ is weak. In this case the standard {   HLR-RPA} theory, almost trivially, gives zero spin hall conductance. The vortex theory, on the other hand, gives a nonzero spin hall conductance due to the Berry phase:
\be
\label{spinhallcond}
\sigma^{spin}_{xy}=-\nu,
\ee
where the unit is taken such that the spin-singlet integer quantum hall effect of fermions at $\nu=2$ has $\sigma^{spin}_{xy}=2$. Notice the interesting minus sign in Eq.~\eqref{spinhallcond}: it implies that the spin hall conductance always has opposite sign with the charge hall conductance.

For $\nu=1$ the above analysis implies that $\sigma^{spin}_{xy}=-1$. This is actually a familiar result in the field theory context: it is simply the ``parity anomaly" of the Dirac composite fermions\cite{parityanomaly0,parityanomaly1,parityanomaly2} in Eq.~\eqref{DCFL}: namely a half-level Chern-Simons term is needed if the $SU(2)$ symmetry is gauged. We will see in Sec.~\ref{PHS} that this is also consistent with the emergent particle-hole symmetry.

\section{Emergent particle-hole symmetry for bosons at $\nu=1$} 
\label{PHS}
 
\subsection{Particle-hole transformation for bosons}
Electronic quantum Hall systems in the lowest Landau level at a filling factor $\nu$ can be viewed in two different ways. We can build them up by starting with an empty Landau level and adding electrons, or by starting with the fully filled Landau level and removing electrons (adding holes). This is known as a particle-hole transformation. At filling factor $\nu = 1/2$, this operation becomes a symmetry (for instance with a 2-body Hamiltonian). As described in previous sections, recent work has described a theory for  the composite fermi liquid at $\nu = 1/2$ that incorporates this symmetry.

In contrast for bosonic quantum Hall systems in the lowest Landau level, the concept of particle-hole transformations apparently makes no microscopic sense. For instance bosons at $\nu = 1$ can form a composite fermi liquid state. Is there an analog of particle-hole symmetry in this state? 

In this section we will see that, indeed there is a reasonable definition of particle-hole transformation for bosons. This definition enables construction of particle-hole conjugates of familiar bosonic quantum Hall states.  In addition at filling $\nu = 1$ the well-developed theory\cite{read98} of the composite fermi liquid is shown to have an emergent particle-hole symmetry. 
At $\nu = 1$ an alternate incompressible non-abelian state - the bosonic Pfaffian - has been studied for a long time. The particle-hole transformation we define enables construction of the particle-hole conjugate of this state which we dub the bosonic anti-Pfaffian as a topologically distinct incompressible state. Depending on the microscopic Hamiltonian, this state may be preferred over the usual bosonic Pfaffian.  We also describe further variations on these bosonic Pfaffian states which follow naturally by pairing the neutral vortex/composite fermions. These are topologically distinct from the standard bosonic Pfaffian or the anti-Pfaffian we introduce below.

For fermions the particle-hole transformation interchanges the $\nu = 0$ state with the $\nu = 1$ integer quantum hall state. Both the empty and filled Landau levels are ``vacua" with trivial excitations and hence can be interchanged by a symmetry. For bosons, apart from the empty vacuum ($\nu = 0)$, there are other interesting vacuua with trivial excitations: the bosonic Integer Quantum Hall (bIQHE)  states which have been discussed recently\cite{luav12,tsml13}. For microscopic models, see Refs.~\cite{sfmu13,wujkj13,nrts13,hesb15,stncnr15,zzdonna16}. These states have electrical Hall conductivity $\sigma_{xy} = 2n$ ($n$ = integer), and thermal Hall conductivity $\kappa_{xy} = 0$. The simplest such state (with $\sigma_{xy} = 2, \kappa_{xy} = 0$) can occur for some interactions at boson filling factor $\nu = 2$. 

Consider now a particle-hole transformation that interchanges bosons at $\nu = 0$ with the bIQHE ground state at $\nu = 2$. It is natural that under this transformation a generic filling factor $\nu$ will go to $2- \nu$. Using this we can define particle-hole conjugates of bosonic quantum hall states: thus the Laughlin states at filling $\nu = \frac{1}{2m}$ lead to a sequence of states at filling $\nu = 2 - \frac{1}{2m}$. 

To understand physically what this transformation describes, consider the bIQHE state at $\nu = 2$. Now imagine doping holes into the system ({\em i.e}, by removing the microscopic bosons). As the excitations of the bIQHE state are just the physical bosons themselves, the holes will be bosons with opposite electric charge.  If the hole filling is $\nu_h$ they can form a Laughlin state at $\nu_h = \frac{1}{2m}$. This leads to the particle-hole conjugate of the usual Laughlin state of particles. 

We can readily write down a wave function that implements the particle-hole transformation.  Let $\psi_p(z_1,.....z_N)$ be a lowest Landau level  wave function of $N$ bosonic particles at coordinates $z_1,.....z_N$ where the total number of flux quanta is $N_\phi$.  As usual the particle filling factor is $\nu_p = \frac{N}{N_\phi}$. The bIQHE occurs at $\nu = 2$. With $N_\phi$ flux quanta this requires an additional $M = 2N_\phi - N$ particles. Let $\psi_{bIQH}(z_1, .......,z_{N+M})$ be the ground state wave function  of the bIQHE state. 
Then the particle hole transformed version of $\psi_p$ is constructed as 
\begin{equation}
\label{phwavefn}
\psi_h([w_i])   =   {\cal N} \int \Pi_{i = 1}^N  d^2z_i \psi_{bIQH}([z_i], [w_i]) \psi_p^*({z_i}).
\end{equation}
(The prefactor ${\cal N}$ on the right is a normalization constant). Here the $w_i$ are $M$ coordinates representing the holes and $z_i$ are $N$ coordinates of particles. The hole filling factor is clearly $\nu_h = \frac{M}{N_\phi} = 2 - \nu_p$ as expected. Note that this is the precise analog of particle-hole conjugation of fermion wave functions.

 \subsection{Bosons at/near $\nu = 1$}
 The particle-hole transformation maps the filling $\nu = 1$ to itself. If $\nu = 1+x$, then under $\mathcal{C}$, $x \rightarrow -x$.  At $\nu = 1$ a compressible composite Fermi liquid (CFL)  state becomes possible. In addition an incompressible non-abelian state (the bosonic Pfaffian) is also possible and is obtained from the composite fermi liquid through pairing. 
 
 It is conceivable that the low energy effective field theory of the CFL has an emergent $\mathcal{C}$ symmetry such that for $\nu = 1+x$, $x \rightarrow - x$.  The standard HLR-RPA theory does not have such a symmetry. Now let us search for such a symmetry in the dual vortex theory of the CFL developed by Read.  The composite fermion $\psi_v$ in this theory should be viewed as an electrically neutral fermion that carries $2\pi$ vorticity. The effective Lagrangian takes the form
 \begin{eqnarray}
 \label{dualvbcfl}
 {\cal L} & =  & \bar{\psi}_v \left(i \partial_t - \mu -a_0 -  \frac{(\vec \nabla - i \vec a)^2}{2m_v} \right) \psi_v \nonumber \\
 && - \frac{1}{2\pi} \epsilon_{\mu \nu \lambda} A_\mu \partial_\nu a_\lambda + \frac{1}{4\pi}\epsilon_{\mu \nu \lambda} A_\mu \partial_\nu A_\lambda.
 \end{eqnarray}
This form of the effective theory was never explicitly written down in Ref.~\cite{read98}; however it can be inferred from the results in that paper. An alternate derivation of this action was provided by Alicea et. al\cite{avl1} using a duality transformation to vortices, and attaching flux to the vortices to convert them to fermions. This theory was also studied recently\cite{mulliganraghu} in the context of field-driven superconductor-insulator transition.
 
The action above in Eq.~\eqref{dualvbcfl} is equivalent to Eq.~\eqref{vortex}, in which the $-2\pi$ Berry phase is dynamically equivalent to zero. Now define an anti-unitary $\mathcal{C}$ operation under which 
 \begin{equation}
 C \psi_v C^{-1} = \psi_v.
 \end{equation}
 This implies that the boson density $\rho_v = \bar{\psi}_v \psi_v$ is even under $\mathcal{C}$. If we now choose the transformations
 \begin{eqnarray}
 Ca_0 C^{-1} & = & a_0, \nonumber \\
 Ca_i C^{-1} & = & - a_i,
 \end{eqnarray}
 we see that $\mathcal{C}$ is a symmetry of the Lagrangian as written. Thus if the low energy physics of a microscopic boson system at Landau level filling $\nu = 1$ is described by this Lagrangian then it has an emergent $\mathcal{C}$ symmetry. It is readily seen that terms violating this particle-hole symmetry all involve higher derivatives of the gauge field and hence  should be   irrelevant\footnote{This is verifiably true in situations where the low energy fixed point is a Fermi liquid as happens when the microscopic interaction is at least as  long ranged as $\frac{1}{r}$ or a weak non-Fermi liquid as happens with  $\frac{1}{r^{1+\epsilon}}$ interactions when $\epsilon$ is small. For short ranged  interactions we do not have good control over the fixed point but a controlled expansion can be developed by taking $\epsilon$ small\cite{nayak94,mrossde}. However only gauge field configurations with small momentum $q$ are expected to be important and so higher derivative terms involving the gauge field  are expected to not affect the ultimate infrared behavior}. For example it was shown in Ref.~\cite{avl1} that the leading $\mathcal{C}$-violating term takes the form $(\nabla\times\vec{a})\cdot(\nabla\times\nabla\times\vec{a})$ and is expected to be irrelevant. The possibility of an emergent particle-hole symmetry for the theory in Eq.~\eqref{dualvbcfl} was raised first in Ref.~\cite{msgmp15}.\footnote{ Ref.~\cite{msgmp15} used the Hamiltonian formalism and an approximate treatment which naturally suggests the emergence of an antiunitary particle-hole symmetry for bosons at $\nu = 1$.  This formalism also gives an explicit representation for the physical charge density in terms of composite fermion degrees of freedom valid at all length scales.} Note that if we vary with respect to  $A_0$ we get 
 \begin{eqnarray}
 \vec \nabla \times \vec a & = &  B - 2\pi \rho \nonumber \\
 & = & B(1-\nu).
 \end{eqnarray}
 
 Thus under $\mathcal{C}$, as $a_i$ is odd, we have $1- \nu \rightarrow \nu - 1$ exactly as expected of a particle/hole symmetry. Exactly at $\nu = 1$ the density operator $\rho - \frac{B}{2\pi}$ is odd under $\mathcal{C}$ again as expected. 
 
Notice that $\kappa_{xy}=0$ for Read's theory, as discussed in Sec.~\ref{thermalhall}. This can be viewed as a consequence of the emergent particle-hole symmetry, since the boson integer quantum hall state at $\nu=2$ has $\kappa_{xy}=0$.
 

\subsection{Bosonic Jain sequence}

As for fermions near $\nu=1/2$, one can access a sequence of incompressible states for bosons near $\nu=1$ by putting the composite fermions into filled Landau levels. In the standard HLR theory, this gives the bosonic Jain sequence\cite{nrtj03}
\be
\nu^{Jain}_{HLR}=\frac{p}{p+1},
\ee
 where $p$ is an integer denoting number of filled CF Landau levels. Notice $p=-1$ appears strange because it actually corresponds to a superfluid. The particle-hole conjugate of the Jain sequence is then
 \be
 \nu^{Jain}_{anti-HLR}=2-\nu^{Jain}_{HLR}=\frac{p+2}{p+1},
 \ee
 {which corresponds to shifting $p$ to $-p-2$. Some of these Jain states with ``negative flux attachment" has been studied numerically in Ref.~\cite{negativep}.}
 
 In Read's theory, the two sequences are united and takes the form
 \be
 \nu^{Jain}_{Read}=\frac{p-1}{p},
 \ee
 where the values with positive $p$ gives the HLR sequence and those with negative $p$ gives the anti-HLR sequence. The superfluid phase here corresponds to $p=0$.

\subsection{Bosonic Pfaffian-like states} 
{  It is extremely interesting to consider paired states of composite fermions that correspond to topologically ordered incompressible fractional quantum Hall states of bosons at $\nu = 1$. It is well known that, if we start with the HLR description of the composite fermi liquid, $p_x-ip_y$ ({\em i.e l = -1} pairing gives the standard bosonic  Pfaffian (Moore-Read) state\cite{readgrn}. This is a non-abelian quantum Hall state with three quasiparticles denoted $1$, $f$ and $\sigma_P$.  Physically the $f$ corresponds to the Bogoliubov quasiparticle associated with the paired state and is an electrically neutral fermion. The $\sigma_P$ is a non-abelion and corresponds physically to the $\pi$ vortex of the ``pair condensate". It carries physical electric charge $q_{\sigma_P} = 1/2$.  The fusion rules are given by
\bea
f\times f &\sim& 1  \nonumber \\
f\times \sigma_P &\sim& \sigma_P \nonumber \\
\sigma_P \times\sigma_P &\sim& b+bf,
\eea
where $b$ is the physical charge-$1$ boson. The topological spin of the Ising-like anyon $\sigma_P$: 
\be
\theta(\sigma_P)=e^{\frac{3i\pi}{8}},
\ee
gives the Abelian part of the braiding statistics of $\sigma_P$. This also leads to thermal Hall conductance $\kappa_{xy}\sim c_-=3/2$, where $c_-$ is the chiral central charge\footnote{Formally the topological quantum field theory corresponding to this familiar bosonic Pfaffian state may be written as $Ising \times U_4(1)/Z_2$.}.

Given the particle-hole conjugation defined in this paper we can clearly construct an alternate bosonic anti-Pfaffian state by starting with the bIQHE state and forming a Pfaffian state of holes with filling factor $\nu_{hole} = 1$. Similar construction of the anti-Pfaffian state has been studied previously\cite{levinapf,ssletalapf} for electrons at $\nu=1/2$. A wave function for the bosonic anti-Pfaffian state may be constructed from that of the standard bosonic Pfaffian using  Eq.~\eqref{phwavefn}. This anti-Pfaffian state has the same three quasiparticles $1, \sigma_{AP}, f$ and charge assignments, and fusion rules. We have changed the subscript of $\sigma$ to $AP$ to denote that it is the non-abelion of the anti-Pfaffian state. The topological spin of $\sigma_{AP}$ is different from the Pfaffian state: 
\be
\theta(\sigma_{AP}) =  e^{\frac{3i\pi}{8}},
\ee
Correspondingly the thermal Hall conductance $\kappa_{xy} \sim c_- = - 3/2$. Thus just like their fermionic counterparts the bosonic Pfaffian and anti-Pfaffian are topologically distinct paired states.  A different route to access the bosonic anti-Pfaffian is to start with the HLR CFL and pair the composite fermions in an $l = 5$ angular momentum channel.  It is readily seen that this leads to the same topological order as the one just described. 

We described these paired states starting with the HLR description of the composite fermi liquid. But what if we start instead with Read's theory Eq.~\eqref{dualvbcfl}? We now show that $l = -1$ pairing of the neutral vortex composite fermions actually leads to a state that is distinct both from the standard bosonic Pfaffian as well as the bosonic anti-Pfaffian introduced above.  It is actually convenient to discuss the general case of pairing with angular momentum $l$ of these neutral vortex composite fermions.  This will clarify the general structure of these various paired states, and establish their connections to those obtained from the HLR construction. 

As the composite fermions in the present theory are single component the pairing is necessarily in an odd angular momentum channel. This then necessarily breaks the particle-hole symmetry. \footnote{This does not mean that particle-hole symmetric gapped state is impossible at $\nu=1$ (see the end of Sec.~\ref{SEG}). It does mean, however, that the particle-hole symmetric state cannot be accessed through a weak pairing starting from Read's theory. Strong pairing can give a particlel-hole symmetric gapped state. } As is well known\cite{readgrn}, such a pairing leads to a non-Abelian topological order in the weak-pairing regime. The details of the topological order, however, depends on the angular momentum of the pairing channel $l=-k$, where $k$ is an odd integer. We denote these topological orders as $k$-Pfaffian states.

Details of the $k$-Pfaffian states ($k$ odd) are as follows: there are two topologically distinct nontrivial quasi-particles, hence three degenerate ground states on a torus. We denote the two quasi-particles as $f$ and $\sigma$, where $f$ represents the composite fermion $\psi_v$ and is therefore a fermion, and $\sigma$ represents $\pi$-vortex in the paired state, and is non-Abelian due to Majorana zero modes associated with the vortex. Since the physical charge is carried by the gauge flux, $f$ carries no physical charge and $\sigma$ carries charge $q_{\sigma}=1/2$. The fusion rules are given by
\bea
f\times f &\sim& 1  \nonumber \\
f\times \sigma &\sim& \sigma \nonumber \\
\sigma\times\sigma &\sim& b+bf,
\eea
where $b$ is the physical charge-$1$ boson. The only data that depends on $k$ is the topological spin of the Ising-like anyon $\sigma$: 
\be
\theta_{\sigma}=e^{\frac{i\pi}{8}k},
\ee
which gives the Abelian part of the braiding statistics of $\sigma$ (see, for example, Ref.~\cite{Kitaev06} for more details). This also leads to thermal Hall conductance $\kappa_{xy}\sim c_-=k/2$, where $c_-$ is the chiral central charge. The familiar bosonic Pfaffian state from HLR theory with $p_x-ip_y$ pairing ($l=-k=-1$) corresponds to $l=-k=-3$ in Read's theory. The shift by $-2$ is a further manifestation of the $-2\pi$ Berry phase associated with the neutral composite fermions that we have discussed in this paper. More generally, a Berry phase $m\pi$ shifts the topological index $k$ (mod $16$) of a superconductor\cite{Kitaev06} from $k=-l$ to $k=-l+m$. For Dirac fermion with $m=\pm1$ this is a well-known property. 

Thus as promised  $p_x - ip_y$ pairing ($l=- 1$) in Read's theory gives a state different from the usual (HLR) bosonic Pfaffian state\footnote{Formally the topological quantum field theory corresponding to this state may be written $\overline{Ising} \times U_4(1)/Z_2$.}.  
 }

It is natural to ask which of these $k$-Pfaffian states will be favorable for a simple bosonic system at $\nu=1$, for example with contact interaction? This of course can only be determined by numerical work. There is an interesting scenario one may anticipate: if the system has approximate particle-hole symmetry, for example when it is close to Read's state in parameter space, then $k$-Pfaffian and $(-k)$-Pfaffian states will be competitive in energy. For example, it is interesting to determine whether bosonic anti-Pfaffian state, which is the particle-hole conjugate (Eq.~\eqref{phwavefn}) of the familiar Pfaffian state, could be energetically competitive in some parameter regime.






\subsection{Particle-hole symmetry for spin-$1/2$ bosons}
The particle-hole symmetry has an interesting twist when the microscopic bosons carry spin-$1/2$, with full $SU(2)$ symmetry.  When spin-$1/2$ bosons are placed in a strong magnetic field (without Zeeman term) they can form a number of quantum Hall states. Of particular importance to us is the fate of such a boson system when the total Landau level filling factor $\nu_{tot} = \nu_\uparrow + \nu_\downarrow = 2$.  Then it is known that (with for instance a simple delta function repulsion interaction) that the result is a boson integer quantum Hall state. Further this state is singlet under the pseudo-spin $SU(2)$ symmetry. Its edge state has counter propagating charge and spin modes. The bulk is characterized by $\sigma_{xy}=2$, $\kappa_{xy}=0$ and $\sigma^{spin}_{xy}=-2$. As in the previous section we may now use this spin singlet bIQHE state to define particle-hole conjugates at other fillings $0 \leq \nu_{tot} \leq 2$. 

We henceforth turn our attention to the special filling $\nu_{tot} = 1$ where states with an emergent particle-hole symmetry might be allowed. It is then clear that the Dirac composite fermi liquid in Eq.~\eqref{DCFL} is such a state. The emergent particle-hole symmetry acts on the composite fermions like time-reversal:
\be
\mathcal{C}: \psi\to\sigma^2\tau^2\psi,
\ee
where $\sigma^i$ denotes Pauli matrices in spin space. Such a form is required by the algebraic structure of the symmetry group $U(1)\times SU(2)\times\mathcal{C}$, namely that $\mathcal{C}$ should commute with $SU(2)$ rotations. Notice $\mathcal{C}^2=1$, in contrary to the Dirac composite fermi liquid state for fermions at $\nu=1/2$. 

As discussed in Sec.~\ref{spinhall}, the spin hall conductance for the above state is $\sigma^{spin}_{xy}=-1$. This is also a consequence of the emergent particle-hole symmetry, since the singlet integer quantum hall state has $\sigma^{spin}_{xy}=-2$ and the particle-hole symmetric state should have half of its spin hall conductance.




 \section{Relation to 3d boson topological insulators}  
 \label{3dSPT}
 
 If $\mathcal{C}$ is indeed an emergent symmetry for the bosonic CFL at $\nu = 1$, we can ask whether the bosonic CFL can be realized in a microscopic system in which $\mathcal{C}$ is an exact local UV symmetry (with in addition the $U(1)$ of charge conservation so that the full symmetry is $U(1) \times \mathcal{C}$). This question was answered for the fermionic CFL at $\nu=1/2$ and lead to fruitful results\cite{sonphcfl}: the fermionic CFL, with the particle-hole symmetry being exact, can be realized on the surface of a $3D$ topological insulator with $U(1)\times\mathcal{C}$ symmetry.
 
 It is then natural to consider bosonic analogs of topological insulator -- also known as symmetry-protected topological (SPT) states\cite{chencoho2011} -- in three dimensions with $U(1) \times \mathcal{C}$ symmetry.
 A simple example\cite{avts12,hmodl,statwitt} is a state that has a response to an external electromagnetic field characterized by a $\theta$ term with $\theta = 2\pi$. In addition the elementary probe magnetic monopole in this electromagnetic field is a neutral fermion that is a Kramers singlet under the anti-unitary $\mathcal{C}$ operation. 
 
 The surface of this boson SPT may be in one of several phases. An example is a $\mathcal{C}$-broken phase without any topological order. There are two states related by the $\mathcal{C}$ symmetry which have electrical Hall conductivity $\sigma_{xy} = \pm 1$, and thermal Hall conductivity $\kappa_{xy} = 0$. Note that these differ precisely by the elementary $2d$ bIQHE state. 
 A different surface state breaks the global $U(1)$ symmetry (a surface superfluid) but preserves $\mathcal{C}$. In this superfluid the basic $2\pi$ vortex is a Kramers singlet fermion while the $4\pi$ vortex is a trivial boson. In addition to these symmetry broken phases, symmetry preserving gapped topological ordered phase with anomalous symmetry implementation are possible. 
 
 As in the analogous discussion of fermions at $\nu = \frac{1}{2}$ in terms of 3d TI surfaces, with $U(1) \times \mathcal{C}$ symmetry, external $B$-fields are $\mathcal{C}$-even and can be included in the Hamiltonian. Then as discussed in Ref. \cite{avts12}, we can get a gapless metallic state by starting from the superfluid surface state and proliferating $2\pi$ vortices which are at a finite density $\frac{B}{2\pi}$. The resulting vortex metal state has exactly the same effective Lagrangian  as Eq.~\eqref{dualvbcfl} but now arises in a system with microscopic $U(1) \times C$ symmetry. 
 
Similar analysis applies to spin-$1/2$ bosons with $U(1)\times SU(2)\times \mathcal{C}$ symmetry: there is a bosonic topological insulator in $3D$ with this symmetry, on the surface of which a natural compressible state is described by the Dirac composite fermi liquid in Eq.~\eqref{DCFL}, without the background Chern-Simons term for $A_{\mu}$ due to the exact particle-hole symmetry. The defining characteristics of this insulator is again a bulk $\theta$-term with $\theta=2\pi$, which makes the elementary magnetic monopole fermionic, with spin-$1/2$ under $SU(2)$ and $\mathcal{C}^2=1$ under particle-hole. The surface avatar of this monopole -- the vortex -- is nothing but the composite fermion.
We describe some more details below, in connection with an amusing electromagnetic duality in the $3d$ bulk.

\subsection{Connection to bulk electromagnatic duality}

The fermionic version of the Dirac CFL at $\nu=1/2$ is deeply connected with a duality between two fermionic topological insulators in three space dimensions, one protected by $\mathcal{C}$ symmetry and the other protected by time-reversal symmetry\cite{tsymmu1,dualdrcwts2015,dualdrmaxav,bulkdualmax}. We now show that the bosonic version is also connected with a duality in three dimensions.

Consider a three dimensional fermion system with $U(1)$ charge conservation, time-reversal $\mathcal{T}$ and $SU(2)$ spin-rotation symmetries, compactly denoted as $(U(1)\rtimes \mathcal{T})\times SU(2)$ (the $\rtimes$ symbol simply means that the $U(1)$ charge is even under $\T$). We also have $\T^2=1$, which is different from the usual cases with fermions. 

With these symmetries, there is a nontrivial topological insulator state for the fermions. The simplest surface state has two Dirac fermion, one for each spin:
\be
\LL[\psi_{\alpha}]=\sum_{\alpha=\uparrow,\downarrow}\bar{\psi}_{\alpha}(i\slashed{\partial}+\mu\gamma^0)\psi_{\alpha},
\ee
which is very similar to Eq.~\eqref{DCFL} except there is no gauge field. Again we have $\T: \psi\to\sigma^2\tau^2\psi$ with $\T^2=1$.

Now let us ``gauge" the entire topological insulator (bulk and surface), by coupling the fermions to a dynamical compact $U(1)$ gauge field. The non-triviality of the topological insulator can be exposed by studying monopoles of the $U(1)$ gauge field. A monopole should have the same quantum number with a $2\pi$-flux tube on the surface, since one can tunnel a monopole from the vacuum into the bulk, which leaves a $2\pi$-flux tube on the surface. As we discussed under Eq.~\eqref{DCFL}, a $2\pi$-flux tube is a spin-$1/2$ boson on the surface. Therefore the monopole in the bulk is also a spin-$1/2$ boson. The monopole charge (magnetic flux) itself is of course odd under time-reversal.

Now the entire three-dimensional $U(1)$ gauge theory can be viewed from a very different angle: one can start from a spin-$1/2$ boson system with $U(1)\times\mathcal{T}\times SU(2)$ symmetry, notice here the first $\times$ symbol means that the $U(1)$ charge is now odd under $\T$. One can then put it into a ``bosonic topological insulator" (BTI), such that when the $U(1)$ symmetry is gauged, the ``monopole" becomes a spin-$1/2$ fermion with $\T^2=1$. This is obviously the same theory as the one above: one only needs to switch the names of ``charge" and ``monopole".

It is conceptually very simple to construct such a BTI, following the recipe introduced in Ref.~\cite{dualdrmaxav}: we can start from the $U(1)$ gauge theory constructed from gauging the fermion TI, then introduce physical spin-$1/2$ bosons with $U(1)\times\mathcal{T}\times SU(2)$ symmetry into the system, and initially make them trivially gapped. We can then hybridize the physical boson with the monopole in the $U(1)$ gauge theory: $\<b^{\dagger}_{\alpha}\mathcal{M}_{\alpha}\>\neq0$, where $b_{\alpha}$ is the physical boson and $\mathcal{M}_{\alpha}$ is the spin-$1/2$ monopole. Because the symmetry quantum numbers of the two particles are identical, such a mixing does not break any symmetry. But it does break the $U(1)$ gauge symmetry and makes the gauge field gapped (and the fermion charge confined). We are thus left with a non-fractionalized insulator in a system of spin-$1/2$ bosons, which is exactly the BTI we are interested in -- the easiest way to see it is to re-introduce a monopole, which will automatically be a spin-$1/2$ fermion.

\section{Topological order: a no-go constraint}
\label{SEG}

It is interesting to ask whether there is a gapped (incompressible) topological order at filling $\nu=1$ that preserves both $SU(2)$ and $\mathcal{C}$ symmetries. For fermions at $\nu=1/2$, such a state (known as PH-Pfaffian) can be obtained by pair-condensing the Dirac composite fermions. If we try to do similar things here, we realize that a spin and particle-hole symmetric pairing term cannot fully gap out the Dirac composite fermions. In fact, we can show that a fully gapped topological order with spin and particle-hole symmetries cannot exist at $\nu=1$. This means that spin-$1/2$ bosons at $\nu=1$ with the full $U(1)\times SU(2)\times\mathcal{C}$ symmetry must be gapless. Such ``symmetry-enforced gaplessness" was first discussed\cite{3dfSPT2} for the surface states of certain fermion topological superconductor. Our example here is a purely  bosonic realization of this phenomenon.

We now prove the statement: suppose there is a fully gapped topological order at $\nu=1$ with particle-hole and $SU(2)$ spin-rotation symmetry. We label the quasi-particles as $\{1,X_1,X_2,...\}$. The physical boson $b_{\alpha}$ is topologically trivial since it has no mutual braiding phase with any particle. So one can freely relabel quasi-particles by attaching certain numbers of $b_{\alpha}$ without changing their topological sector. Now because the spin $SU(2)$ group does not admit fractional (projective) representation (there is no such a thing as ``spin-$1/4$"), each quasi-particle must carry either integer or half-integer spin. We can then relabel the quasi-particles by binding them with certain number of microscopic physical bosons, and make all the topological quasi-particles spin-singlet, which we denote as $\{1,X_1',X_2',...\}$.

Now the topological order $\{1,X_1',X_2',...\}$ (which is the same as the original topological order) is a purely spin-singlet state. In particular, any local objects -- those with trivial mutual statistics with other particles -- in this theory must also be spin-singlet. But we know that spin-singlet local objects in the system are bosons with even-integer charge, because local bosons with odd-integer charge all carry half-integer spin. Therefore the topological order $\{1,X_1',X_2',...\}$ can also be realized in a different system: a system of spinless bosons with even-integer charge.  

It is powerful to view this state from the standpoint of the three dimensional boson SPT for which it is a surface state.  The observations in the previous paragraph imply that the bulk boson SPT can be understood as an SPT of the spin-singlet even charge sector formed out of the elementary bosons. SPT states of spineless bosons with the $U(1) \times C$ symmetry are well understood. It is known that non-trivial such states when protected by the full $U(1) \times C$ symmetry can be characterized by the non-triviality of the elementary magnetic monopole which carries flux $g = \frac{2\pi}{q_e}$ where $q_e$ is the elementary charge of the bosons. But  in the present problem the boson SPTs in question are formed out of the even charge spin-singlet sector which means that we must take $q_e = 2$. However then the monopole with flux $\pi$ is not actually allowed as a legal probe in this system as we also have microscopic charge-$1$ bosons. We do have legal strength-$2\pi$ monopoles but in the even charge boson SPTs these are known to be always trivial. 

However we already know that in our bulk boson SPT the $2\pi$ monopole is non-trivial: it is the bulk avatar of the composite fermion, and hence is a spin-$1/2$ fermion. We have thus reached a contradiction. It follows therefore that the surface of this boson SPT cannot be in a symmetry preserving gapped surface topological ordered state.  This is bosonic example of the phenomenon of ``Symmetry enforced gaplessness", first discovered in the context of fermionic SPT phases\cite{3dfSPT2}. 

Notice that our conclusion relies strongly on the existence of $SU(2)$ invariance. For example, if the entire $SU(2)$ symmetry is absent, there is a $U(1)\times\mathcal{C}$ invariant topological order allowed on the surface of the boson topological insulator, which is then also allowed as a $\mathcal{C}$-invariant topological order at $\nu=1$. It is a simple $\mathbb{Z}_2$ topological order $\{1,e,m,\epsilon\}$, where the two nontrivial bosons $e$ and $m$ both carry half-integer charge. This is  known as the $eCmC$ state in the literature\cite{avts12,hmodl,statwitt}.

\section{Discussion}

We have discussed a candidate theory for compressible states arising from partially filled lowest Landau level. The key ingredients that differ from the traditional HLR-RPA theory are the absence of Chern-Simons term for the emergent gauge field, and a Berry phase on the emergent fermi surface. Our arguments were based on microscopic charge-neutrality of composite fermions in LLL.  {   The composite fermi liquid should thus be viewed as a quantum liquid of neutral fermionic vortices}. We showed that the Berry phase has some direct consequences for transport phenomena in the composite fermi liquids. 

Our picture suggests that for the $\nu=1/2$ CFL state, the $\pi$-Berry phase should be robust even in the absence of particle-hole symmetry, as long as Landau level mixing is suppressed. However in the absence of particle-hole symmetry, it is hard to identify composite fermion density with microscopic operator, and consequently it is hard to detect such a Berry phase through correlation functions at $2k_f$, as was done in Ref.~\cite{geraedtsnum}. It will be desirable to understand the proposed theory from a  microscopic point of view. For example, could there be a microscopic symmetry in the lowest Landau level, that forbids a Chern-Simons term in the effective theory? Such questions are natural directions for future work.

We also described a particle-hole transformation for bosons that relies on removing particles from the boson integer quantum Hall state. This leads to the possibility of a bosonic anti-Pfaffian state at filling $\nu = 1$.  Further $p_x \pm i p_y$ pairing of the neutral vortex/composite fermions  leads to distinct topologically ordered states from the standard bosonic Pfaffian or the anti-Pfaffian we described.  {  This distinction is a manifestation of the $-2\pi$ Fermi surface Berry phase in the neutral vortex liquid theory of the compressible `normal' state of bosons at $\nu = 1$.  We showed that the existing LLL theory for this compressible state   has an emergent particle-hole symmetry.} We emphasized the close parallels between this theory and the particle-hole symmetric composite fermi liquid of fermions at $\nu = 1/2$. 

A different aspect of our paper is our discussion of compressible states of $SU(2)$ symmetric two-component bosons, and their relationship with surface states of three dimensional bosonic topological insulators. In this context we showed that such a boson system does not admit a gapped symmetry preserving state, thereby realizing the phenomenon of ``symmetry enforced gaplessness" discussed in our previous work for fermionic topological superconductors. A general understanding of which $3+1$-D SPT states admit gapped symmetry preserving surface states is another target for future work. 

During the completion of this work, we became aware of another work by Haldane\cite{Haldaneunpub}, through APS March meeting 2016 where part of this work was also presented, that also proposes a Berry phase on the composite fermi surface.  The Berry phase proposed by Haldane apparently differs from ours by a sign. The exact relation between the two works remains unclear.
 
{\textbf{Note added}:  Since the submission of the initial version of this paper, another related paper\cite{mross2pi} has appeared on the arxiv, which also discussed composite fermi liquid states for bosons at $\nu=1$, including one with $2\pi$-Berry phase on the composite fermi surface.}
 
\textbf{Acknowledgement}:  We thank N. R. Cooper, B. I. Halperin, J. K. Jain, O. I. Motrunich, J. Checkelsky, M. A. Metlitski, A. C. Potter, N. Read, and A. Vishwanath for helpful discussions on various issues. Part of this work was performed when TS was visiting the  Aspen Center for Physics,  which is supported by National Science Foundation grant PHY-1066293. CW was supported by the Harvard Society of Fellows. TS was supported by NSF DMR-1305741.  This work was also partially supported by a Simons Investigator award from the
Simons Foundation to Senthil Todadri.

\appendix
\section{Fermi surface Berry phase and quantum oscillations}
\label{fsb}

We consider quantum oscillations, more specifically SdH oscillation, of a fermi surface with Berry phase $\phi_B$. A widely quoted result from Ref.~\cite{Roth, Mikitik} states that the resistivity minima occurs at
\be
\label{A1}
\frac{B_F}{B}=n-\frac{\phi_B}{2\pi},
\ee
where $B_F=2\pi\rho$ and is proportional to the Fermi sea area, and $n$ is some integer. This result was obtained by considering semiclassical (Sommerfeld) quantization of electron orbits on the fermi surface, taking into account the effect of the Berry phase $\phi_B$ when electron moves along the fermi sea. 

However, the above result relies on the assumption that the classical approximation of the ``Fermi surface area", proportional to $B_F$ in Eq.~\eqref{A1}, is independent of $B$. This will not be true if the system has an anomalous Hall conductance. For an ordinary Fermi surface with Berry phase $\phi_B$, one expects an anomalous Hall conductance $\sigma_{xy}=\frac{\phi_B}{2\pi}$ up to an integer. In the presence of this anomalous Hall conductance, part of the electric charge becomes the ``anomalous charge" when the $B$-field is turned on, with density $\rho_A=\sigma_{xy}B/2\pi=\frac{\phi_BB}{4\pi^2}$. This ``anomalous charge" no longer contributes to the Fermi surface area in the Sommerfeld quantization. Therefore the coefficient $B_F$ in Eq.~\eqref{A1} should be modified to 
\be
B_F= 2\pi(\rho-\rho_A)=2\pi\rho-\frac{\phi_B}{2\pi}B
\ee
which exactly cancels the $\phi_B$ term on the right hand side. The final form becomes identical to that of an ordinary Fermi sea without a Berry phase. This is a sensible result since for strong $B$-field, one expects the SdH oscillation to crossover to the integer quantum hall effect, where the integer quantization has no modification from $\phi_B$.

Therefore what the shift in the SdH oscillation really measures is the mismatch between Fermi surface Berry phase and the anomalous Hall conductance. In special circumstances the Berry phase on the Fermi surface does not lead to an anomalous Hall conductance. One of them is the fermionic vortex theory discussed in the main text. Another more familiar example is Dirac fermions in graphene or on topological insulator (TI) surface. For a TI surface the missing Hall conductance is a bulk effect, while for graphene it is due to the cancellation between the four Dirac fermions.

\bibliography{phbsnsbibl}

\end{document}